\documentclass{mem}
\usepackage{natbib}
\usepackage{txfonts}
\usepackage{balance}
\usepackage{graphicx}
\idline{75}{1}
\begin{document}
\def\teff{$T\rm_{eff }$}
\def\kms{km s$^{-1}$}
\def\vv{$V$}
\def\bb{$B$}
\def\ii{$I$}
\def\uu{$U$}
\def\jj{$J$}

\def\umv{$U-V$}
\def\bmv{$B-V$}
\def\umi{$U-I$}
\def\umb{$U-B$}
\def\bmi{$B-I$}
\def\jmk{$J-K$}
\def\cubi{$C_{UBI}$}

\title{Stellar populations in the ELT perspective}


\author{
G.\, Bono\inst{1,2}
\and V.F.\,Braga\inst{1}
\and M.\,Fabrizio\inst{3} 
\and R.\,Gilmozzi\inst{4}
\and R.\,Buonanno\inst{1,3}
\and I.\,Ferraro\inst{2}
\and G.\,Iannicola\inst{2}
\and M.\,Monelli\inst{5}
\and A.\,Milone\inst{6}
\and M.\,Nonino\inst{7}
\and L.\,Pulone\inst{2}
\and P.B.\,Stetson\inst{8}
\and F.\,Thev\'{e}nin\inst{9}
\and A.R.\,Walker\inst{10}
}


\institute{
Univ. Rome "Tor Vergata", Via della Ricerca Scientifica, 1 -- 00133, Roma, Italy
\and
INAF--OAR, via Frascati 33 -- 00040, Monte Porzio Catone (RM), Italy
\and
INAF--OATe, via M. Maggini -- 64100, Teramo, Italy
\and
ESO, Karl-Schwarzschild-Stra{\ss}e 2, 85748, Garching, Germany
\and
IAC, Calle Via Lactea, E38200 La Laguna, Tenerife, Spain
\and
RSAA, The Australian National University, Cotter Road, Weston, ACT, 2611, Australia
\and 
INAF--OAT, via G.B. Tiepolo 11 -- 40131, Trieste, Italy
\and 
DAO--HIA, NRC, 5071 West Saanich Road, Victoria, BC V9E 2E7, Canada
\and 
Obs. C\^{o}te d'Azur, BP 4229 -- 06304, Nice, France
\and 
NOAO--CTIO, Casilla 603, La Serena, Chile
}

\authorrunning{Bono et al.}

\titlerunning{EELT}

\abstract{We discuss the impact that the next generation of  
Extremely Large Telescopes will have on the open astrophysical problems 
of resolved stellar populations. In particular, we address the interplay 
between multiband photometry and spectroscopy.  
 
}
\maketitle{}

\section{Introduction}

The Astronomical Community (AC) is planning the science of  three Extremely Large 
Telescopes (ELTs): Giant Magellan Telescope
(GMT)\footnote{\tt www.gmto.org}, Thirty Meter Telescope
(TMT)\footnote{\tt www.tmt.org} and the European Extremely Large Telescope
(E-ELT)\footnote{\tt www.eso.org/public/teles-instr/e-elt/}. 
These facilities will have a relevant impact on the cutting-edge technology adopted
to build large ground-based telescopes. The transition from the 8-10m class
telescopes to the 30-40m class telescopes before being a "stylistic departure"
is a "paradigm shift" in their realization. The reasons are manifold. 

{\em Adaptive Optics}-- The ELTs plan to have either at first light or during
 operations different flavors of Adaptive Optics systems: ranging from Ground
Layer Adaptive Optics (GLAO) to Single Conjugate Adaptive Optics (SCAO) to Multi
Conjugate Adaptive Optics (MCAO) and to Multi-Object Adaptive Optics (MOAO).
These challenging instruments shaked off their pioneering laboratory phase and
are now producing very interesting results in many astrophysical fields. The
Multiconjugate Adaptive optics Demonstrator (MAD, \citealt{marchetti06})
available at VLT provided a new view of the central regions of Globular Clusters (GCs) 
\citep{ferraro09,bono10mad}, of massive stars \citep{crowther10} and star forming
regions \citep{selman99}. The First Light Adaptive Optics (FLAO,
\citealt{esposito12}) systems available at LBT provided very high quality images
not only of extra-solar planets \citep{esposito13}, but also for accretion disks
\citep{rodigas12} and on the NIR Color-Magnitude Diagram (CMD) of GCs  
\citep{monelli13cubi}. More recently, the adaptive optics system developed at
GEMINI using both natural guide stars and lasers, the Gemini Multi-Conjugate Adaptive
Optics System (GEMS, \citealt{rigaut12}), provided the opportunity to investigate
GCs (NGC~288), open clusters (NGC~2362) and nearby galaxies 
(\citealt{neichel13}; AO4ELT3\footnote{\tt http://ao4elt3.sciencesconf.org/}). 

{\em New theoretical and empirical framework}-- The new facilities will provide
a wealth of new photometric and spectroscopic data. Once again the reduction and
the analysis of these data will require a profound change in the strategy and in
the algorithms currently adopted to perform PSF photometry (DAOPHOT,
\citealt{stetson94}; ROMAFOT, \citealt{buonanno83}; Starfinder,
\citealt{diolaiti00}) and to measure equivalent widths (Fitline,
\citealt{francois03}) or in spectrum synthesis (MOOG, \citealt{sneden73}; ATLAS,
\citealt{sbordone04}) abundance analysis. The same  applies to the new
techniques for integral field spectroscopy that have been developed to deal with
spectroscopic data cubes collected with Integral Field Units (SAURON,
\citealt{jeong12,kamann13}). Theoretical
predictions (stellar evolutionary models, pulsation models, atmosphere models
dynamical models, \citealt{salaris12,bono13}) and cosmological simulations
(Millennium simulation, \citealt{springel05,pandey13}) will also need extensive 
evolution. This applies to 
the physical and numerical assumptions currently adopted to simulate real
phenomena and to the parameter space they are covering.  
  
{\em Science drivers}-- However, the conceptual jump appears even more relevant
if we account for the scientific impact that E-ELT will have on open
astrophysical problems. It is difficult to find a single 
astrophysical field that will not be affected by ELTs. They range from
new constraints on fundamental physic constants \citep{martins13} to measure the
expansion of the universe in real time \citep{liske08} to the reionization of
the universe \citep{pentericci11} to the formation and evolution of galaxies
\citep{puech10} to resolved and unresolved stellar populations \citep{caffau12}
to protoplanetary disks \citep{bast13} and to the characterization of extrasolar
planets \citep{lovis07}.

\section{The link between photometry and spectroscopy} 

In the following we discuss the impact that deep multiband  
photometry and spectroscopy will have on nearby stellar populations.  

\subsection{Age metallicity degeneracy} 

One of the most long-standing and deceptive problems in dealing with 
stellar populations in nearby stellar systems is the so-called 
age-metallicity relation. The key problem is that a stellar population 
can attain redder colors either by an increase in metal content 
\citep{sandage53} or by an increase in age or in 
mean reddening \citep{fanelli88}. The  problem becomes even 
more severe when dealing with unresolved stellar populations \citep{worthey94}.   
The difficulty is a consequence of the fact that broad- and medium-band 
photometry together with spectral indices are affected by an intrinsic 
degeneracy in the case of complex stellar populations \citep{perc_sal09}. 
The observational scenario is further complicated by the evidence that 
these stellar systems are also characterized by a complex star formation
history and probably by changes in the initial mass function. However, 
new indices have been recently suggested to partially 
overcome the degeneracy, and in particular to single out the impact 
of the different astrophysical parameters \citep{vandokkum12}.

Obviously, the age-metallicity degeneracy does not apply to nearby 
resolved stellar populations for which we can clearly identify the 
main-sequence turn-off (MSTO) of the different stellar populations. 
However, even for these systems we still have a problem of age/metallicity 
degeneracy along the Red Giant Branch (RGB). Precise multi-band photometry 
allows us to separate RGB from Asymptotic Giant Branch (AGB) 
stars \citep{stetson11}. However, stellar populations with 
different metallicity distributions and/or different age distributions
are located along the same RGB.  

The typical stellar system in which the above degeneracy is clearly 
present is the dwarf spheroidal galaxy Carina. This galaxy experienced 
at least three different stars formation episodes (see left panel 
of Fig.~1): an old one (t$\sim$12 Gyr) associated with the faint 
subgiant branch (SGB) located at \vv$\sim$23.5 and \bmv$\sim$0.5; an intermediate-age 
one (t$\sim$-6--8 Gyr) associated with the bright SGB located 
at \vv$\sim$22.7 and \bmv$\sim$0.45; and a young one (t$\sim$1 Gyr) 
associated with the bright end of the MS located at \vv$\sim$22.4 and 
\bmv$\sim$0.15.

Data plotted in the left panel of Fig.~1 show that the quoted stellar 
populations do form in the \vv, \bmv\ CMD a very 
well defined RGB in which the width in color is mainly due to 
photometric error (\vv$\sim$21.5, $\sigma_{B-V}\sim$0.02). 
It goes without saying that the use of colors covering a broad range 
in wavelengths typically help to disentangle the different subpopulations. 
However, data plotted in the right panel of Fig.~1 show that even in the 
\vv, \umi\ CMD the width in color along the RGB is still caused by photometric 
errors (\vv$\sim$21.5, $\sigma_{U-I}\sim$0.03). Note that candidate Carina 
stars plotted in the CMDs of Fig.~1 were selected using the \umv, \bmi\ color-color 
plane \citep{bono10car}.

\begin{figure}[ht!]
\includegraphics[trim=.2cm .5cm 1cm 7cm,clip=true,width=1\columnwidth]{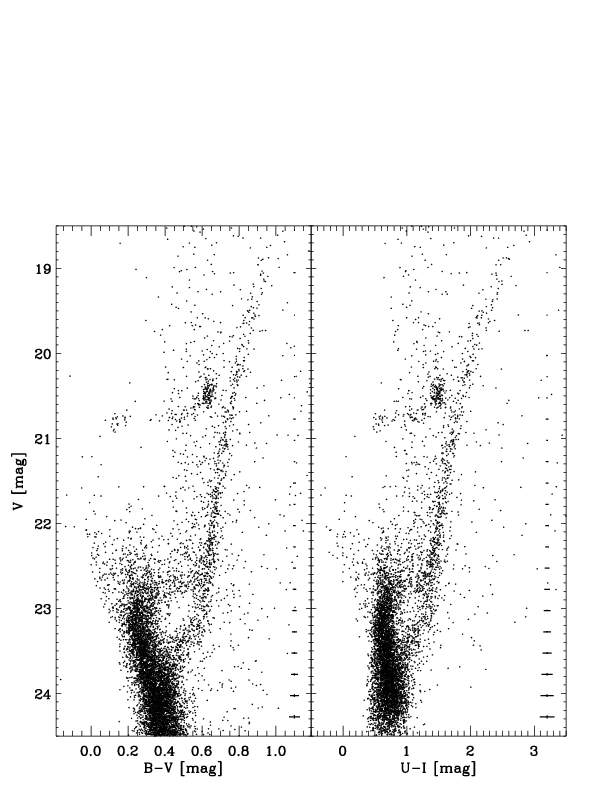}
\caption{\footnotesize Left -- \vv, \bmv\ Color-Magnitude Diagram (CMD) 
of the Carina dSph \citep{bono10car}. Right -- Same as left, but \vv, \umi\ CMD.
The stars plotted in these diagrams were selected by using the \umv, \bmi\
color-color plane.}
\end{figure}

The above empirical scenario also applies to several dwarf galaxies 
in the Local Group (LG). Fig. 2 shows the F814W, F475W--F814W 
CMD of LG dwarfs, based on space images collected with the Advanced Camera 
for Surveys Wide Field Channel on board the Hubble Space Telescope. 
The top panels display the CMD of two isolated dSphs: Tucana (left) and Cetus (right). 
Detailed Star Formation Histories (SFH) of the former system indicate that its
stellar content is dominated by a single, old (12$\pm$1 Gyr) and metal-poor 
SF event \citep{monelli10tuc}. The CMD of the latter isolated dwarf shows 
quite similar features \citep{monelli10cet}. However, Tucana shows two 
well-defined bumps along the RGB, suggestive of double peaked metallicity 
distribution. This empirical evidence is supported by the pulsation properties 
of RR Lyrae stars, showing different period-distributions and different 
mean magnitudes \citep{bernard08}. On the other hand, Cetus shows 
a single RGB bump and its RR Lyrae display homogeneous pulsation 
properties. In spite of the above stark differences the two RGBs 
appear quite similar.

Data plotted in the bottom panels of Fig.~2 show the CMDs of the 
transition galaxy LGS3 and of the dwarf irregular IC1613. The 
morphology of the old, horizontal branch (HB) stars and of the 
intermediate-mass red clump (RC) stars indicates that the quoted 
stellar systems experienced SF activity over a broad range in ages. 
The former galaxy has had an intense SF event from 13.5 to 8.5 Gyr ago \citep{hidalgo11}, 
while IC1613 has had an undergoing SF activity during the last 
12 Gyr. The recent SF event in this system is supported by the well 
populated young main sequence \citep{bernard10} and by the large 
number of classical Cepheids \citep{freedman09}.
Note that the accurate and deep ACS/WFC photometry does not allow us 
to identify the MSTOs of the different subpopulations. 

The above evidence indicates that precise and deep multiband photometry 
for which possible systematic uncertainties in relative and absolute 
photometric zero-points have been constrained at the level of a few 
hundredths of magnitude are still affected by the age/metallicity 
degeneracy along the RGB. This means that current age estimates 
\citep{lemasle12} of stars located along 
the RGB need to be cautiously treated, since current empirical 
and theoretical (mixing length treatment, \citealt{salaris08}) 
uncertainties affect not only the absolute, but also the relative 
age estimates along the RGB.   

\begin{figure}[ht!]
\includegraphics[trim=.5cm 3cm 1.5cm 14.4cm,clip=true,width=1\columnwidth]{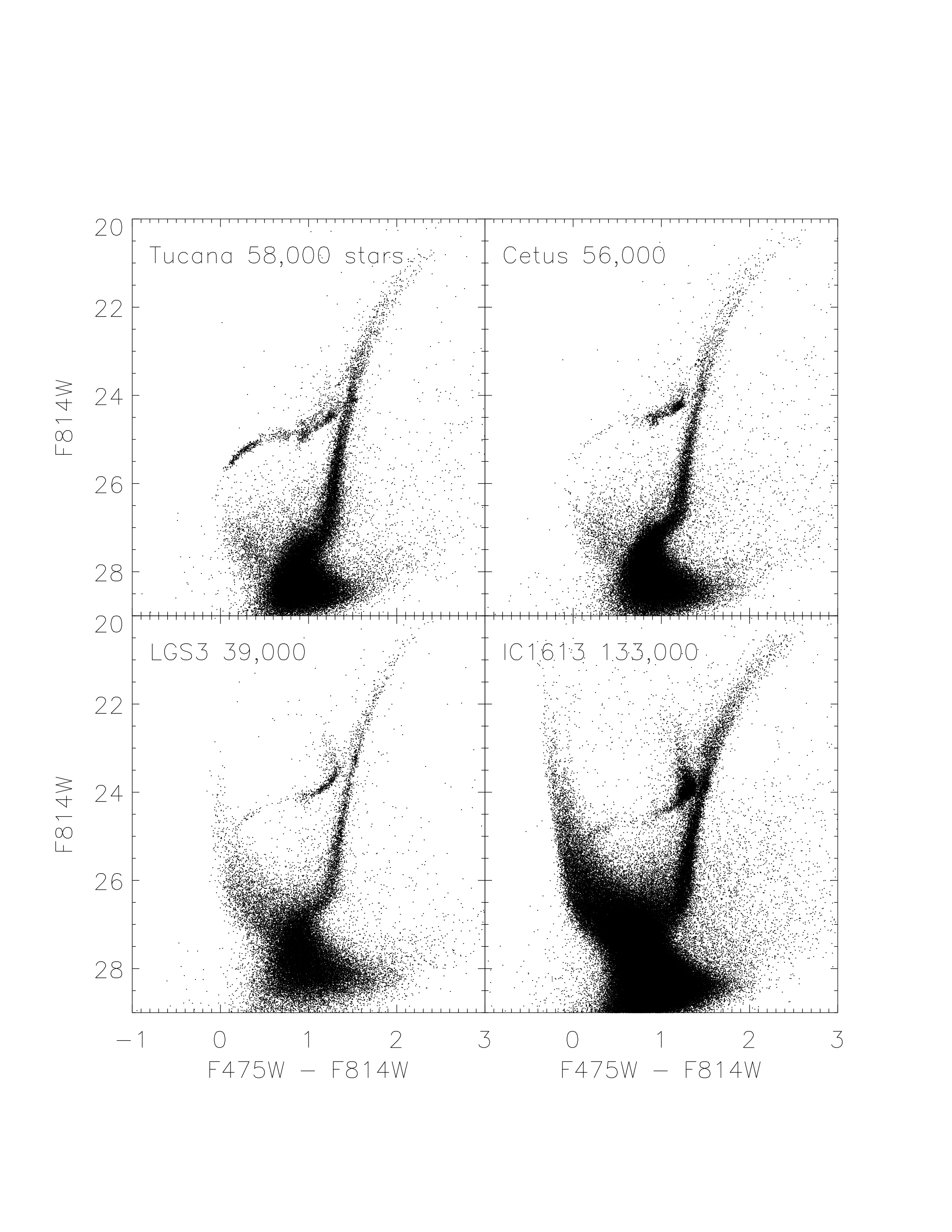}
\caption{\footnotesize 
Top -- F814W, F475W--F814W Color-Magnitude Diagrams (CMD) of two 
Local Group dSphs: Tucana (left) and Cetus (right). The evolved helium burning phase
(F814W$\sim$25, F475W--F814W$\le$1.3 mag) is dominated by old, low-mass HB stars. 
Bottom -- Same as the top, but for the transition dwarf LGS3 (left) and for the 
dwarf irregular IC1613 (right). The helium burning phase includes both HB and 
intermediate-mass, RC stars. The CMD of IC1613 also 
shows a well populated young MS (F814W$\le$25, F475W--F814W$\le$0 mag).}
\end{figure}

\subsection{The \cubi\ index and spectroscopy} 

It came as a surprise that the pseudo-color \cubi=(\umb)--(\bmi) 
introduced by \citet{monelli13cubi} to separate multi-populations in GCs 
can also play a crucial role in disentangling the age-metallicity 
degeneracy in nearby dwarf galaxies. Data plotted in the top left panel of 
Fig.~3 show the \vv, \cubi\ CMD of Carina. Note that the \cubi\ index switches 
hot stars from the left to the right of the CMD. It is worth mentioning that the 
\cubi\ index appears to be quite sensitive to the metal abundance, and indeed 
RGB stars separate into different subgroups. The bluer ones are associated 
with the old population, while the redder ones are associated with the 
intermediate-age population \citep{monelli14car}. This working hypothesis
is supported by a good sample of RGB stars for which 
accurate iron abundances based on high-resolution spectra are available 
\citep{fabrizio12}. The filled and the open circles mark the position of 
the Carina stars associated to the old and to the intermediate-age 
sub-population. Interestingly enough, the same objects do not show any 
significant difference in color in the canonical \vv, \bmv\ CMD (top right 
panel of Fig.~3).  

The bottom panel of Fig.~3 shows the cumulative metallicity distribution 
based on iron abundances provided by \citet{fabrizio12} and by 
\citet{lemasle12} for 68 RGB stars (dashed line). The use of the 
\cubi\ index allow us to constrain the mean metallicity and the 
spread in metallicity of both the old (dark grey area) and the 
intermediate-age (hatched area) subpopulations. We found that 
 the weighted mean metallicity of the old population is --1.81$\pm$0.02,
while its spread is 0.47. On the other hand, the intermediate-age population 
has a mean metallicity of --1.50$\pm$0.01 and a spread of 0.29. Current findings are 
still hampered by the limited sample of spectroscopic measurements, 
but they suggest that the two sub-populations are characterized by 
different metallicity distributions. This preliminary evidence indicates 
that the chemical enrichment history of Carina appears to be much more 
complex than currently assumed.


The use of the \cubi\ index and fiber multi-object spectrographs 
available at the 8m class telescopes will provide the opportunity to 
constrain the chemical enrichment of RGB stars in LG dwarfs 
down to \vv$\sim$21--22 mag. However, the CMDs plotted in Figures 1 and 
3 show that a detailed analysis of the chemical enrichment does 
require large star samples covering the entire body of the nearby galaxies.   

A key advantage of nearby stellar systems is that they can fully exploit 
the intrinsic multiplexity of current and future generations of MOS. 
This means that LG dwarfs are optimal targets for MOS equipped with 
both high-resolution and medium-resolution fibers. The former ones 
can be used for bright RGs to constrain the chemical enrichment of both 
iron and $\alpha$-elements, while the latter to constrain the kinematics.
Note that dwarf galaxies are characterized by a low surface brightness. 
This means that, when moving from the innermost to the outermost regions 
the density of galaxy stars decrease quite rapidly. This means that to 
properly trace the radial velocity profile, and in particular the 
dispersion of the radial velocity, in nearby dwarfs it is mandatory 
to use either SGB or Main Sequence (MS) stars. 

\begin{figure}[]
\includegraphics[trim=.2cm .5cm .5cm .5cm,clip=true,width=1.\columnwidth]{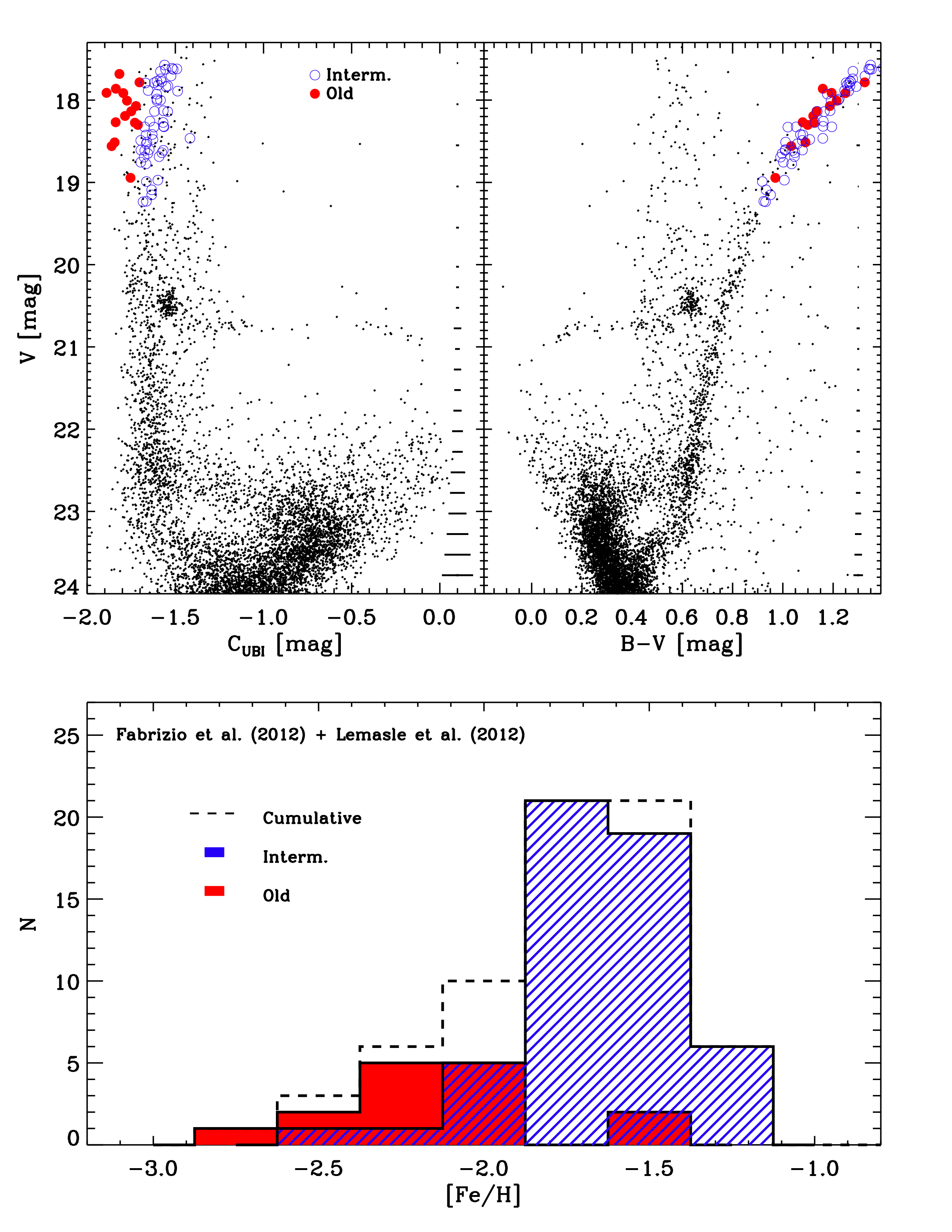}
\caption{\footnotesize Left --  \vv, \cubi\ Color-Magnitude Diagram (CMD) 
of the Carina dSph \citep{bono10car}.
The spectroscopic sample is marked with open triangles \citep{lemasle12} 
and with open circles \citep{fabrizio12}.
Right -- Same as the left, but \vv, \bmv\ CMD. 
Bottom -- Cumulative metallicity distribution (dashed line) according to 
\citet{fabrizio12} and \citet{lemasle12}. The shaded and hatched areas are 
referred to the old and to the intermediate-age subpopulation.}
\end{figure}

The above circumstantial evidence provides solid constraints concerning 
spectral resolution and wavelength coverage for future MOS at ELTs.
An optical, seeing limited, high-resolution (R$\sim$15,000--20,000) spectrograph 
should reach, in a plausible amount of observing time, a limiting magnitude 
of \vv$\approx$24--25 mag. This will enable
high-resolution spectroscopy of RGB and  helium burning stars (HB, RC) 
 over a significant fraction of LG dwarfs. The simultaneous use of 
both low- (R$\sim$1000--2000) and medium-resolution (R$\sim$4000--5000) 
fibers down to a limiting magnitude of \vv$\sim$27--28 means the possibility 
to provide accurate radial velocity measurements for SGB and MS stars 
in the same targets. The need for 3D radial velocity and 
dispersion velocity maps with a precision of $\sim$1--2 \kms\ means 
hundreds of targets per radial bin, and in turn several thousands of 
targets per galaxy. The above requirements suggest a multiplexity between 
low/medium and high-resolution fibers of the order of 400--500.   

\section{Conclusions and final remarks} 

The impact that ELTs will have on astrophysics and cosmology can 
be more easily traced if we account for the context in which 
these new facilities are going to operate. The James Webb Space 
Telescope (JWST) is expected to be launched and operated with 
similar timescales. This means that accurate and very deep 
NIR images of primordial galaxies and galaxy luminosity functions 
as a function of the redshift will become available to the AC. 
Spectroscopic follow up to provide massive estimates of redshifts 
and to use independent diagnostics to constrain the star formation 
rate as a function of redshift together with the characterization of 
their stellar populations and chemical enrichment histories 
appears as sound complements. 

Moreover, we need to take into account that GAIA will start 
taking data quite soon. The photometric, astrometric and spectroscopic 
data that GAIA will collect will provide a new view of the Milky Way 
and of the nearby Universe. The identification of Galactic substructures 
would require deeper multiband photometry and spectroscopy with ELTs.  

On top of the above complementary science ELTs have already a broad 
range of experiments that can only be addressed with ELTs. This 
is the dowry that makes them the most challenging and promising 
opportunity for current and future generation of astronomers.




\begin{thebibliography}{}
\small
\bibitem[Bast et al.(2013)]{bast13} Bast, J.~E., Lahuis, F., van Dishoeck, E.~F., \& Tielens, A.~G.~G.~M.\ 2013, \aap, 551, A118 
\bibitem[Bernard et al.(2008)]{bernard08} Bernard, E.~J., Gallart, C., Monelli, M., et al.\ 2008, \apjl, 678, L21 
\bibitem[Bernard et al.(2010)]{bernard10} Bernard, E.~J., Monelli, M., Gallart, C., et al.\ 2010, \apj, 712, 1259 
\bibitem[Bono et al.(2010)]{bono10car} Bono, G., Stetson, P.~B., Walker, A.~R., et al.\ 2010, \pasp, 122, 651 
\bibitem[Bono et al.(2010)]{bono10mad} Bono, G., Stetson, P.~B., VandenBerg, D.~A., et al.\ 2010, \apjl, 708, L74 
\bibitem[Bono et al.(2013)]{bono13} Bono, G., Inno, L., Matsunaga, N., et al.\ 2013, IAU Symposium, 289, 116 
\bibitem[Buonanno et al.(1983)]{buonanno83} Buonanno, R., Buscema, G., Corsi, C.~E., Ferraro, I., \& Iannicola, G.\ 1983, \aap, 126, 278 
\bibitem[Caffau et al.(2012)]{caffau12} Caffau, E., Bonifacio, P., Fran{\c c}ois, P., et al.\ 2012, \aap, 542, A51 
\bibitem[Crowther et al.(2010)]{crowther10} Crowther, P.~A., Schnurr, O., Hirschi, R., et al.\ 2010, \mnras, 408, 731 
\bibitem[Diolaiti et al.(2000)]{diolaiti00} Diolaiti, E., Bendinelli, O., Bonaccini, D., et al.\ 2000, \aaps, 147, 335 
\bibitem[Esposito et al.(2012)]{esposito12} Esposito, S., Pinna, E., Quir{\'o}s-Pacheco, F., et al.\ 2012, \procspie, 8447,  
\bibitem[Esposito et al.(2013)]{esposito13} Esposito, S., Mesa, D., Skemer, A., et al.\ 2013, \aap, 549, A52 
\bibitem[Fabrizio et al.(2012)]{fabrizio12} Fabrizio, M., Merle, T., Th{\'e}venin, F., et al.\ 2012, \pasp, 124, 519 
\bibitem[Fanelli et al.(1988)]{fanelli88} Fanelli, M.~N., O'Connell, R.~W., Thuan, T.~X.\ 1988, \apj, 334, 665
\bibitem[Ferraro et al.(2009)]{ferraro09} Ferraro, F.~R., Beccari, G., Dalessandro, E., et al.\ 2009, \nat, 462, 1028 
\bibitem[Fran{\c c}ois et al.(2003)]{francois03} Fran{\c c}ois, P., Depagne, E., Hill, V., et al.\ 2003, \aap, 403, 1105 
\bibitem[Freedman et al.(2009)]{freedman09} Freedman, W.~L., Rigby, J., Madore, B.~F., et al.\ 2009, \apj, 695, 996 
\bibitem[Hidalgo et al.(2011)]{hidalgo11} Hidalgo, S.~L., Aparicio, A., Skillman, E., et al.\ 2011, \apj, 730, 14 
\bibitem[Jeong et al.(2012)]{jeong12} Jeong, H., Yi, S.~K., Bureau, M., et al.\ 2012, \mnras, 423, 1921 
\bibitem[Kamann et al.(2013)]{kamann13} Kamann, S., Wisotzki, L., \& Roth, M.~M.\ 2013, \aap, 549, A71 
\bibitem[Lemasle et al.(2012)]{lemasle12} Lemasle, B., Hill, V., Tolstoy, E., et al.\ 2012, \aap, 538, A100 
\bibitem[Liske et al.(2008)]{liske08} Liske, J., Grazian, A., Vanzella, E., et al.\ 2008, \mnras, 386, 1192 
\bibitem[Lovis \& Mayor(2007)]{lovis07} Lovis, C., \& Mayor, M.\ 2007, \aap, 472, 657 
\bibitem[Marchetti et al.(2006)]{marchetti06} Marchetti, E., Brast, R., Delabre, B., et al.\ 2006, \procspie, 6272,  21
\bibitem[Martins et al.(2013)]{martins13} Martins, C.~J.~A.~P., Ferreira, M.~C., Juli{\~a}o, M.~D., et al.\ 2013, arXiv:1309.7758 
\bibitem[Mateo et al.(2012)]{mateo12} Mateo, M., Bailey, J.~I., Crane, J., et al.\ 2012, \procspie, 8446,  
\bibitem[Monelli et al.(2010)]{monelli10cet} Monelli, M., Hidalgo, S.~L., Stetson, P.~B., et al.\ 2010, \apj, 720, 1225 
\bibitem[Monelli et al.(2010)]{monelli10tuc} Monelli, M., Gallart, C., Hidalgo, S.~L., et al.\ 2010, \apj, 722, 1864 
\bibitem[Monelli et al.(2013)]{monelli13cubi} Monelli, M., Milone, A.~P., Stetson, P.~B., et al.\ 2013, \mnras, 431, 2126 
\bibitem[Monelli et al.(2014)]{monelli14car} Monelli, M., et al.\ 2014, \apj, submitted
\bibitem[Neichel et al.(2013)]{neichel13} Neichel, B., D'Orgeville, C., Callingham, J., et al.\ 2013, \mnras, 429, 3522 
\bibitem[Pandey et al.(2013)]{pandey13} Pandey, B., White, S.~D.~M., Springel, V., \& Angulo, R.~E.\ 2013, \mnras, 435, 2968 
\bibitem[Pentericci et al.(2011)]{pentericci11} Pentericci, L., Fontana, A., Vanzella, E., et al.\ 2011, \apj, 743, 132 
\bibitem[Percival \& Salaris(2009)]{perc_sal09} Percival, S.~M., Salaris, M.\ 2009, \apj, 703, 1123 
\bibitem[Puech(2010)]{puech10} Puech, M.\ 2010, \mnras, 406, 535 
\bibitem[Rigaut et al.(2012)]{rigaut12} Rigaut, F., Neichel, B., Boccas, M., et al.\ 2012, \procspie, 8447,  
\bibitem[Rodigas et al.(2012)]{rodigas12} Rodigas, T.~J., Hinz, P.~M., Leisenring, J., et al.\ 2012, \apj, 752, 57 
\bibitem[Salaris \& Cassisi(2008)]{salaris08} Salaris, M., \& Cassisi, S.\ 2008, \aap, 487, 1075 
\bibitem[Salaris(2012)]{salaris12} Salaris, M.\ 2012, \apss, 341, 65 
\bibitem[Sandage(1953)]{sandage53} Sandage, A.~R.\ 1953, \aj, 58, 61 
\bibitem[Sbordone et al.(2004)]{sbordone04} Sbordone, L., Bonifacio, P., Castelli, F., \& Kurucz, R.~L.\ 2004, MemSait, 5, 93 
\bibitem[Selman et al.(1999)]{selman99} Selman, F., Melnick, J., Bosch, G., \& Terlevich, R.\ 1999, \aap, 347, 532 
\bibitem[Smith et al.(2004)]{smith04} Smith, G.~A., Saunders, W., Bridges, T., et al.\ 2004, \procspie, 5492, 410 
\bibitem[Sneden(1973)]{sneden73} Sneden, C.\ 1973, \apj, 184, 839 
\bibitem[Springel et al.(2005)]{springel05} Springel, V., White, S.~D.~M., Jenkins, A., et al.\ 2005, \nat, 435, 629 
\bibitem[Stetson(1994)]{stetson94} Stetson, P.~B.\ 1994, \pasp, 106, 250 
\bibitem[Stetson et al.(2011)]{stetson11} Stetson, P.~B., Monelli, M., Fabrizio, M., et al.\ 2011, The Messenger, 144, 32 
\bibitem[van Dokkum \& Conroy(2012)]{vandokkum12} van Dokkum, P.~G., \& Conroy, C.\ 2012, \apj, 760, 70 
\bibitem[Worthey(1994)]{worthey94} Worthey, G.\ 1994, \apjs, 95, 107 

\end{thebibliography}
\end{document}